\documentclass[useAMS,usenatbib]{mn2e}

\usepackage[T1]{fontenc}
\usepackage{aecompl}
\usepackage{subcaption}
\usepackage{amsmath}
\usepackage{amssymb}
\usepackage{amsmath}
\usepackage{graphicx}
\usepackage{hyperref}
\usepackage{xcolor}
\hypersetup{
    colorlinks,
    linkcolor={red!70!black},
    citecolor={blue!70!black},
    urlcolor={blue!90!black}
}
\newcommand{\diff}{\mathrm{d}}
\usepackage[english]{babel}
\title[Statistical uncertainties and systematic errors]{Statistical uncertainties and systematic errors in weak lensing mass estimates of galaxy clusters}
\author[F. K\"ohlinger et al.]{F. K\"ohlinger$^{1, \, 2}$\thanks{E-mail:
fkoehlin@strw.leidenuniv.nl}, H. Hoekstra$^{1}$, and M. Eriksen$^{1}$\\
$^{1}$Leiden Observatory, Leiden University, P.O. Box 9513, Leiden, NL-2300 RA, the Netherlands \\
$^{2}$Instituut--Lorentz, Leiden University, P.O. Box 9506, Leiden, NL-2300 RA, the Netherlands}

\begin{document}

\date{Accepted 0000 XXXX 00. Received 0000 XXXX 00; in original form 2015 XXXX 00}

\pagerange{\pageref{firstpage}--\pageref{lastpage}} \pubyear{2015}

\maketitle

\label{firstpage}

\begin{abstract}
Upcoming and ongoing large area weak lensing surveys will also discover large samples of galaxy clusters. Accurate and precise masses of galaxy clusters are of major importance for cosmology, for example, in establishing well calibrated observational halo mass functions for comparison with cosmological predictions. We investigate the level of statistical uncertainties and sources of systematic errors expected for weak lensing mass estimates. Future surveys that will cover large areas on the sky, such as \textit{Euclid} or LSST and to lesser extent DES, will provide the largest weak lensing cluster samples with the lowest level of statistical noise regarding ensembles of galaxy clusters. 
However, the expected low level of statistical uncertainties requires us to scrutinize various sources of systematic errors. In particular, we investigate the bias due to cluster member galaxies which are erroneously treated as background source galaxies due to wrongly assigned photometric redshifts. We find that this effect is significant when referring to stacks of galaxy clusters. 
Finally, we study the bias due to miscentring, i.e., the displacement between any observationally defined cluster centre and the true minimum of its gravitational potential. The impact of this bias might be significant with respect to the statistical uncertainties. However, complementary future missions such as eROSITA will allow us to define stringent priors on miscentring parameters which will mitigate this bias significantly.   
\end{abstract}


\section{Introduction}
\label{sec:intro}
Galaxy clusters play an important role in testing cosmological models, for example, by confronting the observed number of galaxy clusters with predictions of the halo mass function (e.g. \citealt{Tinker2008, Tinker2010} and references therein). This test is especially sensitive to the values of the matter content of the Universe, $\Omega_m$, and the normalization of the primordial power spectrum of matter density fluctuations, $\sigma_8$. Values for both parameters obtained from recent Sunyaev-Zel'dovich (SZ) cluster counts by \citet{Planck2014a, Planck2015a} are in tension with other independent measurements (e.g. \citealt{Planck2014b, Planck2015b}). 
To relate the observed cluster counts to predictions for the mass function a conditional scaling relation was used.
The analysis in \citet{Planck2014a} was based on X-ray mass proxies.
The uncertainty in the absolute cluster mass scale remains the largest source of uncertainty in the \textit{Planck} cluster count analyses and is quantified by the mass bias. New priors on this mass bias were incorporated in \citet{Planck2015a} based on small overlapping cluster samples with masses measured by employing gravitational lensing \citep{vdLinden2014, Hoekstra2015}, the deflection of light due to mass as a consequence of Einstein's equivalence principle. Although these improved priors do not fully lift the observed tension yet, scaling relations entirely based on and not only gauged by gravitational lensing measurements are advantageous. This is due to
lensing masses being unaffected by the dynamical state of matter or its physical properties (e.g. being dark or baryonic) in general.   
One disadvantage though is that lensing only yields estimates for a two-dimensional surface mass density, but with simulations it is possible to propagate these reliably into three-dimensional mass estimates in order to compare them with results from other probes (e.g. \citealt{BeckerKravtsov2011, Meneghetti2014}). 

In the strong limit gravitational lensing is characterised by the occurrence of multiply lensed images or arcs of background sources behind the cluster. Employing these lensed images allows a very detailed determination of the mass and mass profile of the cluster core (e.g. \citealt{Medezinski2013}, \citealt{KoehlingerSchmidt2014}; \citealt{Bartelmann2013} and references therein).
In the weak limit small differential deflections of background galaxies are used in a statistical sense to infer the mass of the cluster (e.g. \citealt{Hoekstra2013} and references therein). 

The primary source of statistical uncertainty in weak lensing based cluster mass estimates is shape noise because galaxies are not intrinsically round. For weak lensing it is necessary to measure shapes accurately from observed images in order to derive ellipticity components which serve as shear estimators (cf. Section~\ref{subsec:chi2}). The accumulated errors arising from measuring shapes eventually propagate into weak lensing analyses as shape noise.

Another source of statistical uncertainty is arising from the fact that mass as measured from gravitational lensing is always weighted by the lensing kernel along the line of sight and projected into the plane on the sky of the deflecting mass. The effect of this projected fore- and background mass -- or cosmic noise -- on the accuracy of weak lensing masses has already been studied extensively in the past (e.g. \citealt{Hoekstra2001, Hoekstra2003, Dodelson2004, Hoekstra2011}). 
However and in particular for the cosmological test described above, one is interested in a stack of clusters within a given mass (and redshift) range, thus, statistical uncertainties on properties of the stack will scale inversely with the square root of the total number of clusters within these bins.

In the next decade data from a multitude of ground-based weak lensing surveys, for example, the Kilo-Degree Survey (KiDS\footnote{\url{http://kids.strw.leidenuniv.nl}}, \citealt{deJong2012}), the Dark Energy Survey (DES\footnote{\url{www.darkenergysurvey.org}}, \citealt{Flaugher2005}), and the Subaru Hyper SuprimeCam lensing survey (HSC\footnote{\url{www.naoj.org/Projects/HSC/}}) will become available and eventually culminate in the surveys carried out by the Large Synoptic Survey Telescope (LSST\footnote{\url{www.lsst.org}}, \citealt{Ivezic2008}) and the spaceborne \textit{Euclid}\footnote{\url{www.euclid-ec.org}} mission (\citealt{Laureijs2011}). Although the major focus of these surveys will be cosmic shear -- the much weaker weak lensing due to cosmological large-scale structure --, they will also produce large cluster surveys as 'by-products' and allow mass estimates employing shear measurements (cf. \citealt{Sartoris2015}). Since all of these weak lensing surveys will provide superior statistics in terms of the expected number of clusters to be found due to an orders of magnitude increase in survey area, it is important to scrutinize sources of statistical and systematic uncertainties in the determination of cluster masses using weak gravitational lensing. 

In this paper, we will be especially focussing on the \textit{Euclid} survey because it will yield one of the largest cluster samples with a very low level of statistical uncertainties based on weak lensing. This allows us to scrutinize the impact of major sources of uncertainties and biases, eventually answering whether these large cluster surveys will be more affected by statistical uncertainties or systematic errors. Note, however, that the survey design for LSST is very similar to \textit{Euclid} so that our results will also be applicable to this survey and to lesser extent to smaller surveys such as, for example, DES. 

Due to the large increase in survey area, we expect \textit{large} samples of clusters to be detected in ongoing and upcoming weak lensing surveys and hence the statistical uncertainties are expected to decrease to levels on which systematic errors will no longer be negligible but instead might even dominate over statistical uncertainties.
As we have mentioned already, gravitational lensing always yields a projected mass due to the line of sight integration over the lensing kernel. Preferably, the redshifts of (all) background source galaxies should be used in the integration over the lensing kernel (cf. Section~\ref{subsec:chi2}) which thus requires to take redshifts of even larger samples of galaxies. In the ongoing and upcoming surveys this will be achieved by employing photometric redshift estimates based on multiband observations since spectroscopy is not feasible anymore given the typical survey areas (at least several thousand square degrees). 
   
However, photometric redshift estimates are less precise and accurate and photometric misidentifications, for example, in the cluster member galaxy assignment will propagate into a biased mass estimate via the lensing kernel. Similarly, the effect of miscentring -- the displacement between any observationally defined cluster centre and the minimum of the gravitational potential of the cluster -- can also be propagated into a bias of weak lensing mass estimates. On the scale of a single cluster these biases are negligible but this might not be anymore the case once we turn to studying the masses of large ensembles of clusters.

In this paper, we will determine the level of expected statistical uncertainties first and then continue to study possibly important sources of bias.
Eventually, these biases will have to be assessed much more rigorously through extensive (numerical) simulations, the aim of this paper is, however, to provide a guideline for the design of these simulations by identifying the most significant sources of bias with respect to the expected level of statistical uncertainties for stacks of galaxy clusters.

The structure of the paper is as follows: in Section~\ref{sec:stat_uncert} we establish the level of expected statistical uncertainties on the mass estimates of stacks of galaxy clusters from the \textit{Euclid} survey while describing the weak lensing formalism used in this study at the same time. In Section~\ref{sec:sys_errors} we scrutinize various sources of bias, most importantly the effect of photometric redshift outliers as well as miscentring and compare them to the level of statistical uncertainties derived in the previous section. Finally, we present our conclusions in Section~\ref{sec:conc}. 

Throughout this paper we employ a spatially flat $\Lambda$CDM cosmology with $\Omega_m=0.308$, $\Omega_{\Lambda}=0.692$, $H_{0}=100 h \, \mathrm{km \, s^{-1} \, Mpc^{-1}}$ with $h = 0.678$, $\sigma_8=0.826$ and $n_s=0.961$ following results from \cite{Planck2014b}.

\section{Statistical uncertainties}
\label{sec:stat_uncert}
Estimating the total number of haloes in redshift and cluster mass bins is the first step in answering the question of whether a \textit{Euclid} cluster survey will be limited by statistics or systematics since the statistical uncertainty on a stack of clusters scales inversely with the square root of the total number of haloes in the stack. In a real data analysis one has to stack the clusters according to an observational proxy (e.g. luminosity). This will introduce an Eddington bias \citep{Eddington1913} in the stacked quantity as a function of decreasing number density. Hence, the average halo masses on the high mass end will be lowered. This in turn reduces the signal-to-noise ratio of the weak lensing measurement. However, this would only weaken the constraints on systematic errors we derive below. Therefore, we do not assume a proxy for the stacking but stack the clusters directly in mass in order to derive the tightest constraints on systematic errors.

The \textit{Euclid} survey will cover an area of $15000 \deg^2$ on the sky and is expected to detect $30$ galaxies per square arcminute \citep{Laureijs2011} for which accurate shapes can be determined. For the subsequent analysis we will assume that all clusters within this area will be detected down to redshifts of $z = 1.5$ spanning masses between $6.78 \times 10^{13} h^{-1}\mathrm{M_\odot}$ and $2.70 \times 10^{15} h^{-1}\mathrm{M_\odot}$. This assumption is deliberately optimistic because it leads to the tightest constraints on systematic errors. Note, however, that for example \citet{Gladders2007} have found their high-redshift sample of clusters from the Red-Sequence Cluster Survey (RCS) to be complete to $\approx 88\%$.

\subsection{Halo abundance}
\label{subsec:halo_abundance}
The halo abundance can be expressed in the functional form \citep{Tinker2008, Tinker2010}
\begin{equation}
 n(M, \, z) = \frac{\diff n}{\diff M} = \frac{\bar{\rho}_{m, 0}}{M} f(\nu) \frac{\diff \nu}{\diff M} \, .
\label{eq:HMF}
\end{equation}
The function $f(\nu)$ is motivated from extended Press-Schechter theory and can explicitly be written as \citep{Tinker2008, Tinker2010}
\begin{equation}
 f(\nu) = \alpha[1+(\beta\nu)^{-2\phi}]\nu^{2\eta} \, e^{-\gamma\nu^2/2} \, ,
\end{equation}
where the parameters $\alpha, \beta, \gamma, \eta$, and $\phi$ are redshift-dependent and have to be calibrated against numerical simulations for the corresponding overdensity $\Delta = 200$ with respect to $\bar{\rho}_m$ (cf. \citealt{Tinker2008, Tinker2010} for explicit values of these parameters).

The halo abundance is predicted as a function describing the mass fraction of matter in peaks of a given height, $\nu \equiv \delta_c / \sigma(M, z)$, in the linear density field smoothed at a scale $R = (3M/(4 \pi \bar{\rho}_{m, 0}))^{1/3}$ \citep{PressSchechter1974}, where $M$ and $z$ refer to the cluster mass and redshift, respectively, and $\bar{\rho}_{m, 0}$ is the mean matter density of the Universe today. The constant $\delta_c = 1.686$ denotes the critical overdensity for collapse in linear theory and $\sigma(M, z)$ is the root-mean-square (rms) variance of the linear density field smoothed on a scale $R(M)$, which is defined as
\begin{equation}
 \sigma^2 = \int_0^\infty P(k) |\hat{W}(k \cdot R)|^2 k^2 \, \diff k \, .
\end{equation}
Here, $P(k)$ is the linear matter power spectrum, which we calculate with the fitting formulas provided by \citet{EisensteinHu1999} and $\hat{W}$ is the Fourier transform of the top-hat filter with radius $R$ in real space.
 
Using equation~\ref{eq:HMF} we can now predict the (expected) abundance of haloes per redshift bin $i$ and mass bin $j$ by evaluating
\begin{equation}
\label{eq:abundance}
 N_{ij} = \int_{z_{\mathrm{low}, i}}^{z_{\mathrm{high}, i}} \int_{M_{\mathrm{low}, j}}^{M_{\mathrm{high}, j}} \int_V n(M_j \, , z_i) \, \diff z \, \diff M \,  \diff V_{\mathrm{com}} \, ,
\end{equation} 
where we integrate over the expected comoving volume of the \textit{Euclid} survey. The subscripts 'low' and 'high' refer to the lower and upper bounds of the bin, respectively.

\subsection{Mass model}
\label{subsec:model}
Next we have to specify a mass model for galaxy clusters from which we will derive shears that can then be compared to the measured shear around galaxy clusters. Numerical simulations of cosmological volumes show that the Navarro-Frenk-White (NFW)--profile (\citealt{NavarroFrenkWhite1997, Navarro2010}) is a good description of the average density profile of an ensemble of haloes over several orders of magnitude in mass when adjusting the halo concentration accordingly. 

In the following analysis we assume only a single halo component and a spherically symmetric distribution of the cluster mass. In general, this assumption is over-simplifying and especially for unrelaxed single haloes far from correct (e.g. \citealt{Shaw2006}). However, since we focus in our analysis on a stacked signal from an ensemble of clusters, this simplification holds, because non-spherical symmetric cluster geometries will average out in the stacking process provided that the selection of the cluster sample is unbiased. An unbiased cluster sample is an important assumption here in order to derive upper limits on systematic errors, but in a real data analysis the cluster selection function has to be fully taken into account, e.g., in a subsequent cosmological analysis.

In this case the radial profile of such an idealized halo can then be expressed as an NFW--profile:
\begin{equation} 
\label{eq:NFW}
 \rho(r) = \frac{\delta \cdot \bar{\rho}_{m}(z)}{r/r_s(1+r/r_s)^2} \, ,
\end{equation}
where $\bar{\rho}_{m}(z)$ is the \textit{mean matter density} of the Universe at the redshift $z$ of the halo. The parameter $\delta$ describes the overdensity of the halo and is related to the concentration parameter $c$ through
\begin{equation}
 \delta = \frac{200}{3} \frac{c^3}{\ln(1+c)-c/(1+c)} \, .
\end{equation}  
The scale radius $r_s$ is a characteristic radius of a cluster and can be related to the virial radius $r_{200}$ and concentration parameter $c$ via $r_s = r_{200}/c$. We define the virial radius here as the radius of a sphere which contains a mass overdensity of $200\bar{\rho}_m(z)$. Thus, the corresponding mass $M_{200}$ within this sphere is given by
\begin{equation}
 M_{200} = \frac{800\pi}{3}\bar{\rho}_m(z)r_{200}^3 \, .
\end{equation}
Furthermore, numerical simulations hint at a (noisy) relation between halo concentration and mass. By applying such a concentration-mass-relation, we can reduce the free parameters of the model to only one: the mass $M_{200}$.

For the concentration--mass relation we use the results of \citet{DuttonMaccio2014}, i.e.,
\begin{equation}
 \log_{10}(\hat{c}_{200}(\hat{M}_{200})) = a + b\log_{10}(\hat{M}_{200}/(10^{12} \, h^{-1} \mathrm{M_{\odot}}))\, 
\label{eq:Mc}
\end{equation} 
with the redshift-dependent functions $a=0.520+(0.905-0.520)\exp(-0.617z^{1.21})$ and $b=-0.101+0.026z$. The concentration $\hat{c}_{200}$ and mass $\hat{M}_{200}$ are defined with respect to the \textit{critical density} of the Universe. We convert between this definition and our definition of mass and concentration given with respect to the \textit{mean matter density} employing the algorithm from \citet{HuKravtsov2003}. Note that at an earlier stage of the subsequent analysis we employed the concentration--mass relation from \citet{Duffy2008} which qualitatively did not affect any of our subsequent results or conclusions. That is expected because the weak lensing signal depends to first order on mass only. 
Moreover, we do not assume any scatter in the concentration--mass relation, because scatter will mainly affect the shape of the profile at small scales. However, our analysis always assumes that we measure the weak lensing signal of a stack and that these these small scale fluctuations from halo to halo due to scatter in the concentration--mass relation average out.  

 
\subsection{Weak lensing formalism}
\label{subsec:chi2}
Analytical formulas for the calculation of the weak lensing convergence and shear signal from a spherically symmetric NFW--profile were derived in \citet{Bartelmann1996} and are conveniently re-expressed in \citet{WrightBrainerd2000}. Following these references, we write the convergence as
\begin{equation}
\label{eq:kappa}
\kappa_{\mathrm{NFW}}(x) = \frac{\Sigma_{\mathrm{NFW}}(x)}{\Sigma_{\mathrm{crit}}} \, ,
\end{equation}
which is thus the ratio of the surface density $\Sigma_{\mathrm{NFW}}(x)$ at projected position $x=R/r_s$ scaled by the critical surface density
\begin{equation}
\label{eq:sig_crit}
\Sigma_{\mathrm{crit}} = \frac{c^2}{4\pi G D_l(z_{l})} \, \beta^{-1}(z) \, ,
\end{equation} 
where $\beta(z)^{-1} = D_{s}(z)/D_{ls}(z, z_{l})$ is the inverse of the lensing efficiency $\beta(z)$.
Here $D_s$, $D_l$, and $D_{ls}$ denote the angular diameter distances between observer and source, observer and lens, and lens and source, respectively. The constants $c$ and $G$ are the speed of light and gravitational constant, respectively. For explicit formulas of the surface density $\Sigma_{\mathrm{NFW}}(x)$ of an NFW--profile we refer the reader to the original literature \citep{Bartelmann1996, WrightBrainerd2000}. 

As we have hinted at already in Section~\ref{sec:intro}, weak lensing requires the knowledge of the redshift of the lens and every background source which are entering as variables in the corresponding angular diameter distances. However, instead of considering a redshift for every single background galaxy one assumes or measures a source redshift distribution:
\begin{equation}
p_{\mathrm{src}}(z) = \frac{\beta}{z_0 \Gamma(\frac{1+\alpha}{\beta})}\left(\frac{z}{z_0}\right)^\alpha \exp(-(z/z_0)^\beta) \, , 
\label{eq:pz}
\end{equation}
where we have adopted the functional form presented in \citet{Vafaei2010} and use $\alpha=0.96$, $\beta=1.70$, and $z_0=1.07$ corresponding to a median redshift of $z_{\mathrm{med}}=0.91$ to simulate the \textit{Euclid} survey. Employing this source redshift distribution lets us rewrite the critical surface density as 
\begin{equation}
\label{eq:sig_crit_pz}
\Sigma_{\mathrm{crit}} = \frac{c^2}{4\pi G D_l(z_l)} \int_{z_{\mathrm{min}}}^{z_{\mathrm{max}}} \diff z \, \beta_{\mathrm{eff}}^{-1}(z) \, ,
\end{equation}
with the inverse of an effective lensing efficiency $\beta_{\mathrm{eff}} =  p(z) \cdot \beta(z)$. Note that the source distribution $p(z)$ has now to be renormalized over the range $z_{\mathrm{min}} \leq z \leq z_{\mathrm{max}}$. 

In case of \textit{Euclid} and all other ongoing and upcoming  lensing surveys, photometric redshifts will also be available. Hence, we will only consider galaxies as sources for the lensing signal with redshifts $z_{\mathrm{min}} = z_{\mathrm{phot}} > z_{\mathrm{cluster}} + 0.15$ where we choose an offset of $0.15$ because the lensing contribution of sources close to the cluster redshift is negligible and for low redshifts the offset of $0.15$ corresponds to the expected $3 \sigma$ uncertainty in photometric redshift $\sigma_z = 0.05(1+z)$.

The tangential shear due to an NFW-profile can be expressed as 
\begin{equation}
\gamma_{T}^{\mathrm{NFW}}(x) = \frac{\bar{\Sigma}_{\mathrm{NFW}}(x)-\Sigma_{\mathrm{NFW}}(x)}{\Sigma_{\mathrm{crit}}} \, ,
\end{equation}
i.e., as a scaled density contrast between the average surface density inside projected radius $x$ and the surface density at radius $x$.  
However, observationally, it is only possible to measure the reduced tangential shear $g_{T}$, i.e.,
\begin{equation}
g_{T} = \frac{\gamma_{T}}{1 - \kappa} \, .
\end{equation}
A parametrized model based on the equations above can then be used to derive the mass of the halo from the measured shear signal. Here, we will also include the effect of cosmic noise. This is important in order to derive more realistic uncertainties on the mass estimates \citep{Hoekstra2001, Hoekstra2003, Dodelson2004, Hoekstra2011}. The most straightforward implementation in order to achieve that is fitting the parametric model directly to a pixelized map of the two Cartesian projections of the reduced shear $g_{T}$ in the lens plane. These are related to $g_{T}$ through
\begin{align}
g_1 &= g_{T} \cos(2\phi) \\
g_2 &= g_{T} \sin(2\phi) \, .
\end{align}
For the implementation of cosmic noise contributions we follow \citet{Oguri2010} and calculate the $\chi^2$ as
\begin{equation}
	\begin{split}
 	\chi^2 = \sum_{\alpha, \beta = 1}^2 \sum_{k, l = 1}^{N_{\mathrm{pixel}}} & [g_\alpha(\boldsymbol{\theta_k})-g_\alpha^m(\boldsymbol{\theta_k;p})][\boldsymbol{C^{-1}}]_{\alpha \beta, k l} \\
           & \times [g_\beta(\boldsymbol{\theta_l})-g_\beta^m(\boldsymbol{\theta_l;p})] \, ,
	\end{split}
\label{eq:chi2}
\end{equation}
where Greek indices run over the two components of the reduced shear $g$ and Roman indices run over the pixel positions ($k, l = 1,..., N_{\mathrm{pixel}}$). The matrix $\boldsymbol{C}$ denotes the covariance matrix and $\boldsymbol{C^{-1}}$ is its inverse.

By minimizing the $\chi^2$--value given the distortion 'data', we find the best-fitting model parameters. Of course, with data we are referring to the shear components derived from a fiducial parametric model. 

We consider now two contributions in the covariance matrix: the dominating intrinsic ellipticity noise, i.e., shape noise, and cosmic noise due to large-scale structure along the line of sight. Thus, we write the covariance matrix as
\begin{equation}
\boldsymbol{C} = \boldsymbol{C^{\mathrm{shape}}} + \boldsymbol{C^{\mathrm{LSS}}}.
\label{eq:lss1}
\end{equation}
The intrinsic ellipticity noise between different galaxies is uncorrelated, thus, the shape noise covariance matrix consists only of diagonal terms
\begin{equation}
[\boldsymbol{C^{\mathrm{shape}}}]_{\alpha \beta, k l} = \delta_{\alpha \beta} \delta_{k l} \, \sigma_{\mathrm{shape}}^2 \, ,
\label{eq:lss2}
\end{equation}
where $\delta_{ij}$ denotes the Kronecker delta and $\sigma_{\mathrm{shape}}^2 = \sigma_{\mathrm{int}}^2/N_k$ is the shape noise in pixel $k$. We estimate the number $N_k$ of background galaxies per pixel as:
\begin{equation}
N_k = A_{\mathrm{pix}} \cdot n \int_{z_{\mathrm{min}}}^{z_{\mathrm{max}}} \diff z \, p_{\mathrm{src}}(z) \, ,
\label{eq:number_density}
\end{equation}
where $A_{\mathrm{pix}}$ stands for the area of one pixel and the source redshift distribution $p_{\mathrm{src}}(z)$ is the one given in equation~\ref{eq:pz}. We employ a number density of background sources of $n = 30 \, \mathrm{arcmin^{-2}}$ which is expected for the \textit{Euclid} survey \citep{Laureijs2011}. Furthermore, we assume an intrinsic ellipticity noise per galaxy of $\sigma_{\mathrm{int}} = 0.25$ per component. 

In contrast to that, large-scale structure along the line of sight introduces correlated noise which can be expressed as \citep{Hoekstra2003, Dodelson2004}
\begin{equation}
[\boldsymbol{C^{\mathrm{LSS}}}]_{\alpha \beta, k l} = \xi_{\alpha \beta}(r=|\boldsymbol{\theta_k}-\boldsymbol{\theta_l}|) \, ,
\label{eq:lss3}
\end{equation} 
where $\xi_{\alpha \beta}$ denote the cosmic shear correlation functions. Since the universe is statistically isotropic, we express $\xi$ as a function of the length $r$ of the vector connecting the two positions $\boldsymbol{\theta_k}$ and $\boldsymbol{\theta_l}$. In particular, the shear correlation functions constructed from the two shear components are given by
\begin{align}
\xi_{11}(r) &= \cos^2(2\phi) \xi_{++}(r) + \sin^2(2\phi)\xi_{\times \times}(r) \, , \\
\xi_{22}(r) &= \sin^2(2\phi)\xi_{++}(r) + \cos^2(2\phi) \xi_{\times \times}(r) \, , \\
\xi_{12}(r) &= \xi_{21}(r) = \cos(2\phi)\sin(2\phi)[\xi_{++}(r)-\xi_{\times \times}(r)] \, , 
\end{align}
where $\phi$ is the position angle between the coordinate $x$-axis and the vector $\boldsymbol{r} = \boldsymbol{\theta_k}-\boldsymbol{\theta_l}$. $\xi_{++}$ and $\xi_{\times \times}$ denote the tangential and cross-component shear correlation functions (e.g. \citealt{BartelmannSchneider2001}), respectively. For the calculation of these shear correlation functions we have to calculate the non-linear matter power spectrum $P_\delta(k)$ folding in the source redshift distribution defined in equation~\ref{eq:pz}. We calculate the non-linear matter power spectrum with the publicly available Boltzmann-code \texttt{CLASS\footnote{Version 2.4.0 from \url{www.class-code.net}}} \citep{Blas2011, Audren2011}. 


In all of our subsequent analyses, we fit the shear signal on a regular grid with constant side length $N_{\mathrm{pixel}}=20$ which is set to correspond to a square with side length $2 \times 2 \, R_{\mathrm{vir}}$. Furthermore, we cut out the cluster centre in a square of side length $2 \times 0.2 \, R_{\mathrm{vir}}$ (we will refer to that more conveniently as fitting 'from $0.2 \, R_{\mathrm{vir}}$ to $2 \, R_{\mathrm{vir}}$' from now on). Larger scales than $2 \, R_{\mathrm{vir}}$ are avoided since these are completely dominated by cosmic noise due to the smallness of the cluster signal there. In addition to that, two-halo contributions will also bias the NFW--fit on these scales \citep{BeckerKravtsov2011}. Smaller scales than $0.2 \, R_{\mathrm{vir}}$ are omitted since this is usually done for practical purposes when dealing with real data in order to minimize the residual contamination by cluster members. Furthermore, the accuracy of shape measurements in high density and hence high shear regions is also an issue one tries to circumvent in practice. Finally, we also do expect deviations from the simple NFW--profile on these scales due to effects of substructure \citep{BeckerKravtsov2011}.

\subsection{Results}
\label{subsec:stat_uncert_results}
With the formalism outlined above we can now turn to determining the statistical precision of mass estimates from weak lensing measurements.
Most importantly, we emphasize that realistic estimates of the statistical uncertainties must include cosmic noise (cf. equation~\ref{eq:lss1}).

\begin{figure*}
	\centering
	\begin{minipage}{180mm}
    	\centering
		\begin{subfigure}{0.49\textwidth}
			\centering			
			\includegraphics[width=\textwidth]{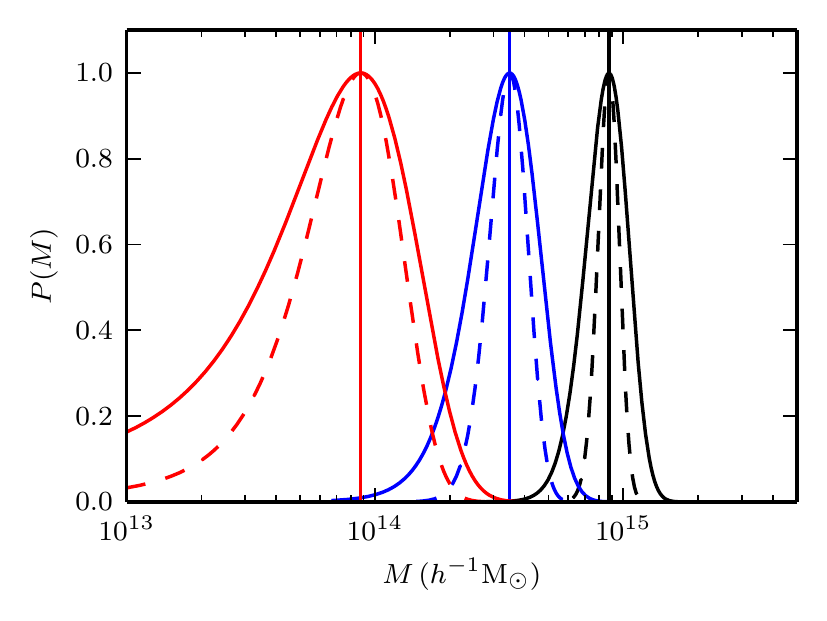}
			\caption{}
			\label{fig:PDF_mass}
		\end{subfigure}
        \begin{subfigure}{0.49\textwidth}
           	\centering
			\includegraphics[width=\textwidth]{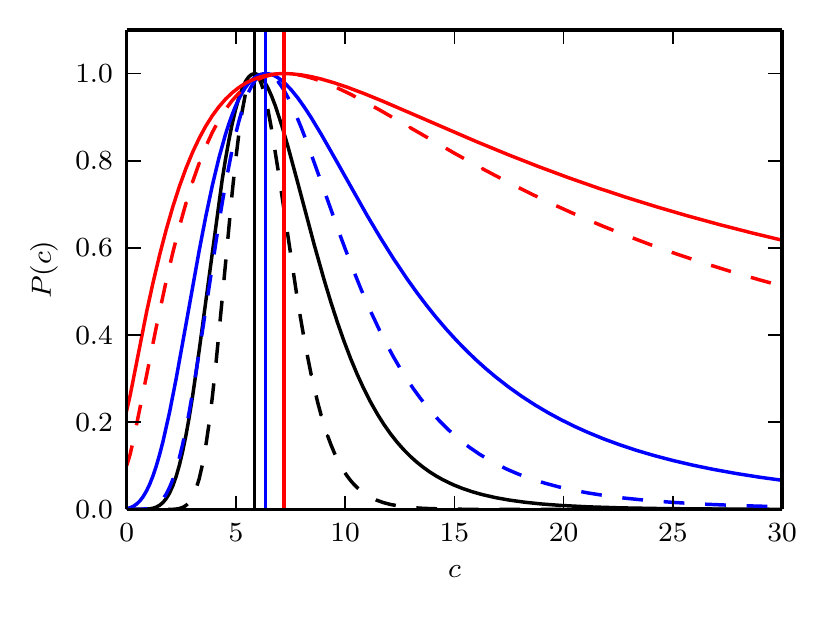}
			\caption{}
			\label{fig:PDF_conc}
		\end{subfigure}
		\caption{\textbf{(a)} Likelihood as a function of halo mass $M$ and concentrations derived through the concentration-mass relation 		of equation~\ref{eq:Mc} for single haloes of masses $M_1 = 0.87 \times 10^{14} \, h^{-1}\mathrm{M_\odot}$ (red), $M_2 = 3.50\times 		10^{14} \, h^{-1}\mathrm{M_\odot}$ (blue) and $M_3 = 0.87 \times 10^{15} \, h^{-1}\mathrm{M_\odot}$ (black). \\
		\textbf{(b)} Likelihood as a function of concentration $c$ at fixed fiducial mass $M$ for the same three haloes as in panel a) (fiducial concentrations are set by employing the concentration-mass relation of equation~\ref{eq:Mc}.\\
		In both panels vertical lines indicate the fiducial values of mass and concentration, respectively. Dashed lines do not include, 			and solid lines do include large-scale structure contributions in the covariance (cf. equation~\ref{eq:lss1}).}
	\end{minipage}
\end{figure*}

This is demonstrated by Fig.~\ref{fig:PDF_mass} and Fig.~\ref{fig:PDF_conc} which show the likelihoods $P(M)$ and $P(c)$ as functions of the halo mass $M$ and concentration $c$, respectively, resulting from the fitting procedure described above for three halo masses in the range $0.87 \leq M/ (10^{14} \, h^{-1} \mathrm{M_\odot}) \leq  8.76$, all at the same redshift $z=0.1875$. In the derivation of $P(M)$ we use the concentration-mass relation given in equation~\ref{eq:Mc}. In the derivation of $P(c)$ the mass was fixed at the fiducial value for each cluster and the fiducial concentration again derived through employing equation~\ref{eq:Mc}. We then fit for the concentration $c$ and demonstrate the importance of accounting for large-scale structure contributions in the data covariance.

The dashed lines show the distributions only taking into account shape noise, whereas for the solid lines the effect of large-scale structure is additionally taken into account (cf. equation~\ref{eq:lss1}, \ref{eq:lss2}, \ref{eq:lss3}). This results in a significant broadening of the corresponding distributions $P(M)$ and $P(c)$, which in turn increases the uncertainties in the halo mass and concentration significantly. This effect is expected because large-scale structure contributions reduce the weight of large scales on the weak lensing signal. Therefore, it is more pronounced for lower mass haloes since there the lensing signal is much weaker on larger scales compared to high mass haloes. 

We continue the analysis by looking further at the stacked weak lensing signal of clusters in eight mass bins from $6.78 \times 10^{13} h^{-1}\mathrm{M_\odot}$ to $2.70 \times 10^{15} h^{-1}\mathrm{M_\odot}$ and in four different redshift bins in the range $0 < z < 1.5$. 

The relative error of the stacked lensing signal of a halo in each redshift bin $i \in {0,...,3}$ and mass bin $j \in {0,...,7}$ is estimated by
\begin{align}
\label{eq:signal}
 \Delta M_{ij}^{\mathrm{left/right}} &= |M(\Delta\chi^2=1)^{\mathrm{left/right}}-M_j(z_i)| \, , \\
 \sigma_{ij} &= \frac{1}{2}\frac{(\Delta M_{ij}^{\mathrm{left}}+\Delta M_{ij}^{\mathrm{right}})}{M_j(z_i)} \, ,
\end{align}
where the $\chi^2$ is calculated again as defined in equation~\ref{eq:chi2}. The minimal $\chi^2$ is centred on the central value of each mass and redshift bin, respectively.

\begin{figure*}
	\centering
	\begin{minipage}{126mm}
		\centering
		\includegraphics[width=\textwidth]{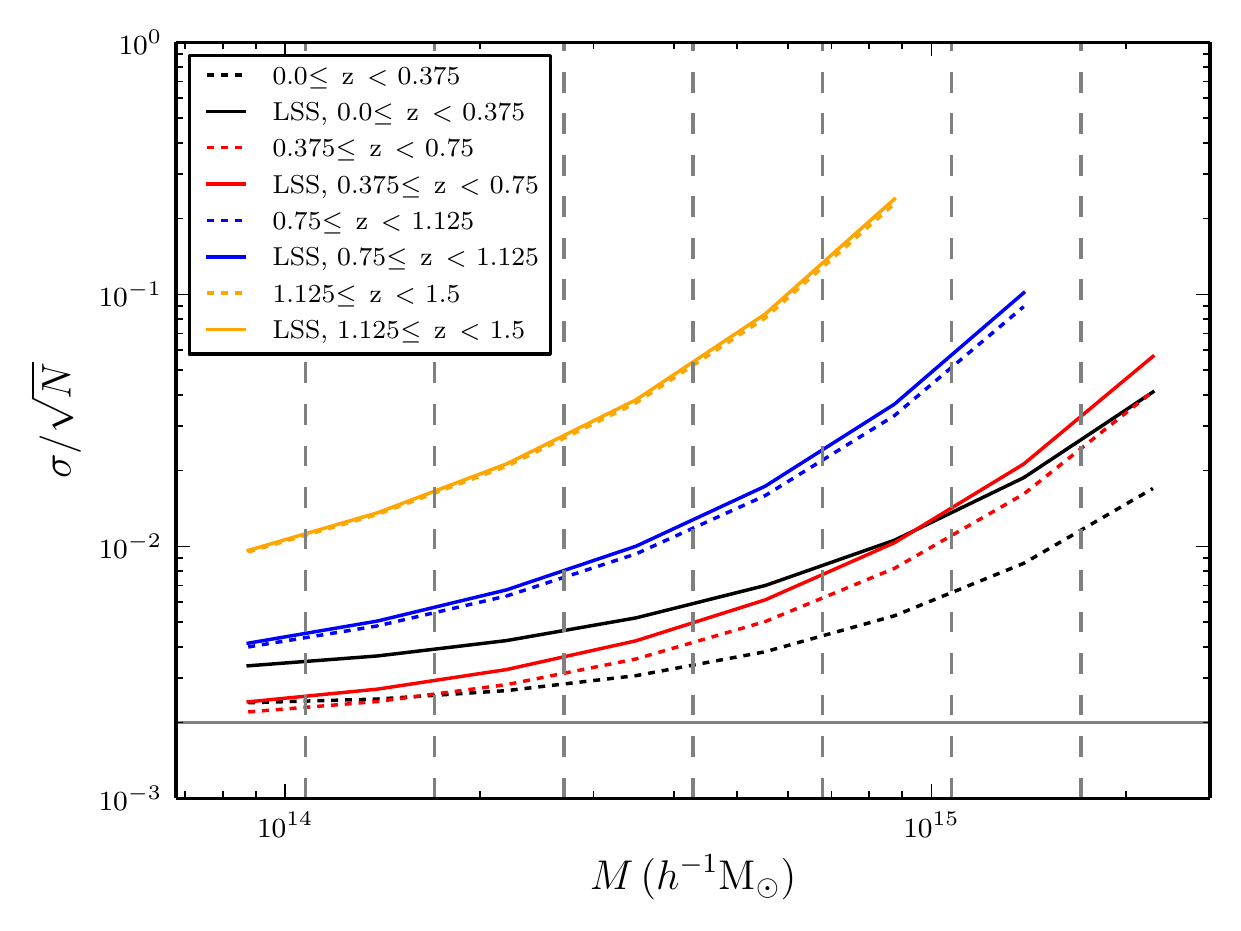}
		\caption{Relative uncertainties for masses of galaxy clusters as estimated from a 
		stacked weak lensing signal for different mass and redshift bins. The solid lines take 
		large-scale structure contributions into account whereas the dashed lines only include 
		shot noise contributions (cf. Section~\ref{subsec:chi2}). The eight mass bins (vertical, 
		dashed lines) span a range from $6.78 \times 10^{13} h^{-1}\mathrm{M_\odot}$ to $2.70 
		\times 10^{15} h^{-1}\mathrm{M_\odot}$ and the four redshift bins are in the range $0 < 
		z < 1.5$. The horizontal grey line indicates the required upper bound on the multiplicative bias $m$ for \textit{Euclid} (cf. Section~\ref{sec:sys_errors}).}
		\label{fig:forecast}
	\end{minipage}
\end{figure*}

In Fig.~\ref{fig:forecast} we show the expected statistical relative uncertainty on the mass for a stacked weak lensing signal from galaxy clusters in each of the mass and redshift bins specified above.
The solid lines correspond to the total relative error including also large-scale structure contributions in the error budget. These are most important for clusters at low redshifts $z \leq 0.75$ and increase with increasing halo mass driven mainly by the lower abundances of these high mass clusters.

In Table~\ref{tab:numbers} we provide the quantitative values for the relative uncertainties in all mass bins for the lowest redshift bin since these ultimately set the requirements on the precision.  
\begin{table*}
\begin{minipage}{126mm}
\caption{Relative uncertainties in mass.}
\label{tab:numbers}
\begin{tabular}{@{}ccccccr}
\hline
  $M_{\mathrm{low}} \, (\times 10^{14} \, h^{-1} \mathrm{M_\odot})$ &
  $M_{\mathrm{high}} \, (\times 10^{14} \, h^{-1} \mathrm{M_\odot})$ &
  $\sigma_{\mathrm{stat}}$ &  
  $\sigma_{\mathrm{stat}}/\sqrt{N}$ &
  $\sigma$ & 
  $\sigma/\sqrt{N}$ &
  $N$ \\
\hline
  0.678 & 1.075 & 0.432 & 0.0024 & 0.605 & 0.0034 & 32509\\
  1.075 & 1.703 & 0.322 & 0.0025 & 0.477 & 0.0037 & 16861\\
  1.703 & 2.699 & 0.239 & 0.0027 & 0.378 & 0.0042 & 7999\\
  2.699 & 4.278 & 0.178 & 0.0031 & 0.302 & 0.0052 & 3363\\
  4.278 & 6.780 & 0.132 & 0.0038 & 0.242 & 0.0070 & 1200\\
  6.780 & 10.75 & 0.098 & 0.0053 & 0.196 & 0.0106 & 342\\
  1.075 & 17.03 & 0.073 & 0.0086 & 0.159 & 0.0187 & 72\\
  17.03 & 26.99 & 0.054 & 0.0170 & 0.130 & 0.0410 & 10\\

\hline
\end{tabular}

\medskip
The relative errors $\sigma/\sqrt{N}$ for each mass bin of Fig.~\ref{fig:forecast} are given for the lowest redshift bin (i.e., $z_{\mathrm{mean}} = 0.1875$). $N$ refers to the number of haloes in each mass bin and $M_{\mathrm{low}}$ and $M_{\mathrm{high}}$ denote the lower and upper value of each mass bin, respectively.
\end{minipage}
\end{table*}

\section{Systematic errors}
\label{sec:sys_errors}
So far, our analysis did not include any sources of bias. However, given the expected small statistical uncertainties as estimated in the previous section (cf. Fig.~\ref{fig:forecast} and Table~\ref{tab:numbers}), any source of systematics might become important. 

In general, the simplest parametrization of systematic deviations of the observed shear $\gamma_{\mathrm{obs}}$ from the true shear $\gamma$ can be written as \citep{Heymans2006}
\begin{equation}
\gamma_{\mathrm{obs}} = (1+m)\gamma+c \, ,
\end{equation} 
where $m$ refers to the multiplicative bias and $c$ to the additive bias (e.g. \citealt{Huterer2006}; \citealt{Massey2013} and references therein). 

The multiplicative bias can arise from a variety of different sources. 
For example, the shear is derived from an observed image which is a convolution of a point spread function (PSF) of finite size with the true shape of the observed object. This convolution introduces a multiplicative error.  
The additive bias, for example, can result from anisotropies of the PSF. We expect these to largely average out when referring to a stack of galaxy clusters which implies taking an average over many cluster--source pairs.   

Since the requirement on the multiplicative bias for \textit{Euclid} is $m < 2 \times 10^{-3}$ \citep{Laureijs2011}, we find that the statistical uncertainties are still larger and thus this source of systematic can be neglected henceforth. Therefore, the primary observational sources of bias in weak lensing can both be neglected for our subsequent analysis in which we will focus then on other possible sources of bias instead.  

\subsection{Photometric redshift bias}
\label{subsec:photo-z}
For all ongoing and upcoming weak lensing surveys photometric redshifts of large samples of galaxies will be available. Equations~\ref{eq:sig_crit} and ~\ref{eq:sig_crit_pz} show the dependence of the cluster lensing signal on redshifts of the cluster itself and all background sources or a source redshift distribution, respectively. Since it would simply require too much time to obtain spectroscopic redshifts for all galaxies in and around each cluster in these large samples of galaxy clusters, photometric redshift estimates present the only feasible alternative. However, the techniques for estimating photometric redshifts from multiband observations are not as precise as spectroscopic redshifts. They might even yield catastrophic outliers, for example, due to misinterpretations or confusion of emission or absorption features (e.g. Ly$\alpha$ break; cf. \citealt{Jouvel2011}).
  
This is also important for the weak lensing signal since cluster member galaxies that are scattered to higher redshifts will now be treated as source background galaxies in the measurement of the weak lensing signal. In comparison to the case of a perfect assignment of cluster members and background source galaxies the weak lensing signal will be diluted because erroneously scattered cluster members are not gravitationally sheared by the cluster. In this context another bias might arise due to cluster members being intrinsically aligned although \citet{Sifon2015} showed no evidence for this. Note that any member galaxy scattered to a redshift in the foreground of the cluster does not change the lensing signal, though, because the photometric redshift distribution will be cut for all redshifts lower than $z_{\mathrm{cut}} = z_{\mathrm{clus}} + 0.15$. 

In order to quantify the level of contamination, we need an estimate of the number of cluster members we expect. For that reason, we adopt a Schechter-type cluster luminosity function \citep{Schechter1976} which follows the scaling relations with respect to cluster mass (or richness) derived from a large sample of groups and clusters from the Sloan Digital Sky Survey (SDSS) as presented in \citet{Hansen2009}. For simplicity we only assume a passive redshift evolution of the luminosity function. The absolute magnitudes sampled from this cluster luminosity function are converted to apparent magnitudes which will then serve as input for an estimation of photometric redshifts for the cluster member galaxies. 
\begin{table}
\caption{The SN = 5 limiting (apparent) magnitudes for extended objects used in the photometric redshift estimation.}
\label{tab:sn}
\begin{center}
\begin{tabular}{ c c c c c c c }
\hline
g& r& i& z& Y& J& H\\
\hline
24.65& 24.15& 24.35& 23.95& 23.3& 23.3& 23.3\\
\hline
\end{tabular}
\end{center}
\medskip 
These limits assume DES ground based observations (g, r, i, z) and \textit{Euclid} NIR observations (Y, J, H). The limiting magnitudes for extended objects are assumed to be 0.7 magnitudes shallower than for point sources.
\end{table}

In general, observed magnitudes include noise from a variety of sources, for example, errors in the background subtraction, CCD readout noise, zodiacal light, and photometric errors. However, in our simulations it is sufficient to only use fixed magnitude limits. In Table~\ref{tab:sn} we show the signal-to-noise (SN) limits for different bands, combining DES optical photometry (e.g. \citealt{Banerji2015}) and \textit{Euclid} near infrared (NIR) observations \citep{Laureijs2011}.
%
%
%
The photometric redshifts are estimated from the apparent magnitudes by employing a template based photo-z code derived from \texttt{BPZ} \citep{Benitez2000}. A single redshift is estimated by finding the peak of
\begin{equation}
\chi^2(z) = \sum_{i,T} \frac{\left(\tilde{f_i} - f_i(z,T)\right)^2}
{\sigma^2_{f_i}} \, ,
\end{equation}
\noindent
where the sum is taken over the different filters $i$ and different types of spectral energy distributions (SED) $T$ for mimicking different types of galaxies. Here $\tilde{f_i}$ is the observed flux in filter $i$, $\sigma_{f_i}$ is the flux error, and $f_i(z,T)$ is a model flux constructed from the template library.
In the estimation of the photometric redshifts, we do not include the \textit{Euclid} visual (VIS) band. 
Furthermore, no priors are included and the templates equal the ones used to generate the simulations.

This yields a catalogue of photometric redshift estimates for cluster members for which we do know the true cluster redshift by construction. 
We cut the catalogue by imposing a detection limit for extended objects of $m^{\mathrm{VIS}} < 24.5 \ (10\sigma)$ in apparent magnitude following the requirement for lensing sources as given in \citet{Laureijs2011}. After having applied this cut we use the catalogue to construct a photometric redshift distribution for the cluster members, $p_{\mathrm{clus}}(z)$, which we use to define the modified total source redshift distribution $p_{\mathrm{mod}}^i(z)$ per radial bin $i$: 
\begin{equation}
\label{eq:p_mod}
p_{\mathrm{mod}}^i(z) \propto N_{\mathrm{src}}^i \cdot p_{\mathrm{src}}(z) + N_{\mathrm{clus}}^i \cdot p_{\mathrm{clus}}(z) \, .
\end{equation}
Here $N_{\mathrm{src}}^i$ denotes the total number of source galaxies per radial bin, i.e., $N_{\mathrm{src}}^i = (1-f_{\mathrm{src}}^{\mathrm{scat}}) \cdot A_{\mathrm{fit}}^i \cdot n$. The number density of sources is again assumed to be $n = 30 \, \mathrm{arcmin}^{-2}$ and with $A_{\mathrm{fit}}^i$ we denote the area of an annulus defined by the borders of the radial bin. Note that the actual number of source galaxies per radial bin entering in the weak lensing analysis is again calculated as in equation~\ref{eq:number_density} with $p_{\mathrm{src}}$ replaced by a properly normalized $p_{\mathrm{mod}}$. 

Moreover, we account for the fact that background source galaxies are also scattered into the cluster foreground by introducing $f_{\mathrm{src}}^{\mathrm{scat}}$, the fraction of background sources scattered into the foreground of the cluster. These additional source scatterers intensify the effect of cluster galaxies scattered into the background twofold: firstly, the number of sources per bin is lowered which increases the uncertainty of the weak lensing signal and secondly the last term of $p_{\mathrm{mod}}^i(z)$ modified by the scattered cluster galaxies gets a higher, relative weight. We estimate $f_{\mathrm{src}}^{\mathrm{scat}}$ from an additionally simulated source redshift distribution employing the photometric redshift estimation algorithm described above but using now luminosity functions from \citet{Marti2014}.

The source redshift distribution has the same functional form as in equation~\ref{eq:pz}. We estimate the number of cluster members in the same radial bin $i$ as 
\begin{equation}
N_{\mathrm{clus}}^i = f_{\mathrm{fit}}^i \cdot N_{\mathrm{clus}}^{\mathrm{tot}}(\leq 2\, R_{\mathrm{vir}}) \, ,
\end{equation}
where $f_{\mathrm{fit}}^i$ is the fraction of cluster members in the radial bin $i$, i.e., 
\begin{equation}
f_{\mathrm{fit}}^i = \frac{\int_{R_{\mathrm{low}}^i}^{R_{\mathrm{high}}^i} \Sigma(R) R \, \mathrm{dR}}{\int_{0}^{2 \, R_{\mathrm{vir}}} \Sigma(R) R \, \mathrm{dR}} \, .
\end{equation}
The surface mass distribution $\Sigma(R)$ is derived from the NFW--profile given in equation~\ref{eq:NFW} (e.g. \citealt{vdBurg2014}). The number of cluster members is estimated by integrating the luminosity function up to the absolute magnitude corresponding to the detection limit in visual apparent magnitude $m^{\mathrm{VIS}}$ multiplied with the volume of a cylinder with radius $2 \, R_{\mathrm{vir}}$ and height $4 \, R_{\mathrm{vir}}$. Note that the calculations as presented above are slightly inconsistent with our approach of effectively fitting within squares enclosing the corresponding annulus. However, as can be seen in Fig.~\ref{fig:forecast_photoz}, the bias decreases with increasing radius and thus this approach does not change our conclusions below.

When we fit masses, we cut and normalize $p_{\mathrm{mod}}(z)$ over the range $z_{\mathrm{cut}} \leq z \leq z_{\mathrm{max}}$, where $z_{\mathrm{max}}$ should be formally set to infinity. We use here, however, a high enough redshift beyond which $p_{\mathrm{mod}}(z) = 0 \ \forall z \geq z_{\mathrm{max}}$. The fiducial model to which we compare in the fitting makes use of the same normalizations in $p_{\mathrm{mod}}(z)$, but with the cluster redshift distribution $p_{\mathrm{clus}}(z)$ set to $0$.

\begin{figure}
		\centering
		\includegraphics[width=84mm]{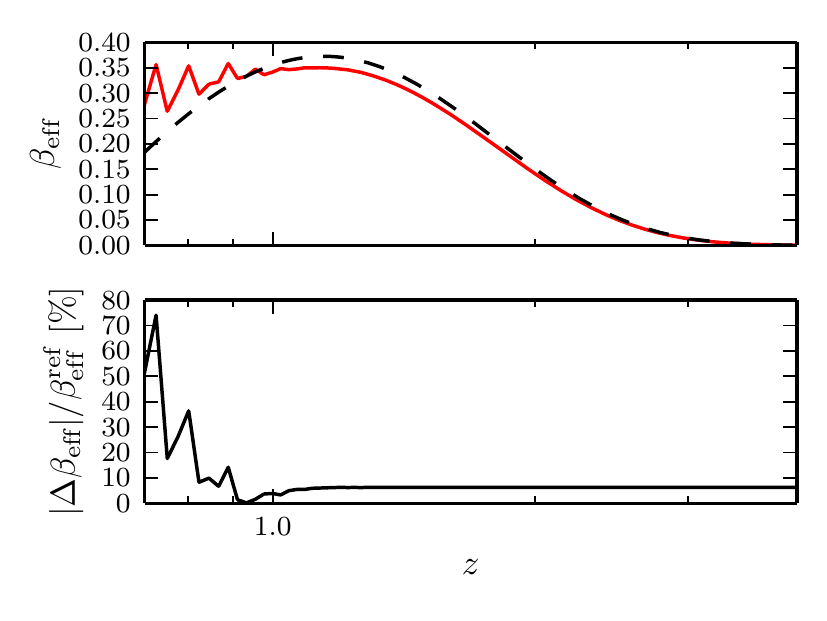}
		\caption{\\ \textbf{Upper panel:} Effective lensing efficiencies $\beta_{\mathrm{eff}}(z)$ for the reference source background distribution (black, dashed line; cf. equation~\ref{eq:pz}) and for a modified source redshift distribution (red, solid line; cf. equation~\ref{eq:p_mod}) for a cluster of mass $M = 8.76 \times 10^{14} \, h^{-1} \, \mathrm{M_\odot}$ at redshift $z_l = 0.56$ in the fitting range of the first radial bin, i.e., $0.2 \, R_{\mathrm{vir}} \leq r \leq 1 \, R_{\mathrm{vir}}$. The modified source redshift distribution consists only of elliptical galaxies (i.e., SED type ``Ell01'').\\
		\textbf{Lower panel:} Relative difference between both effective lensing efficiencies.\\
		}
		\label{fig:sig_crit_pz}
\end{figure}

\begin{figure*}
	\centering
	\begin{minipage}{180mm}
    	\centering
		\begin{subfigure}{0.49\textwidth}
			\centering			
			\includegraphics[width=\textwidth]{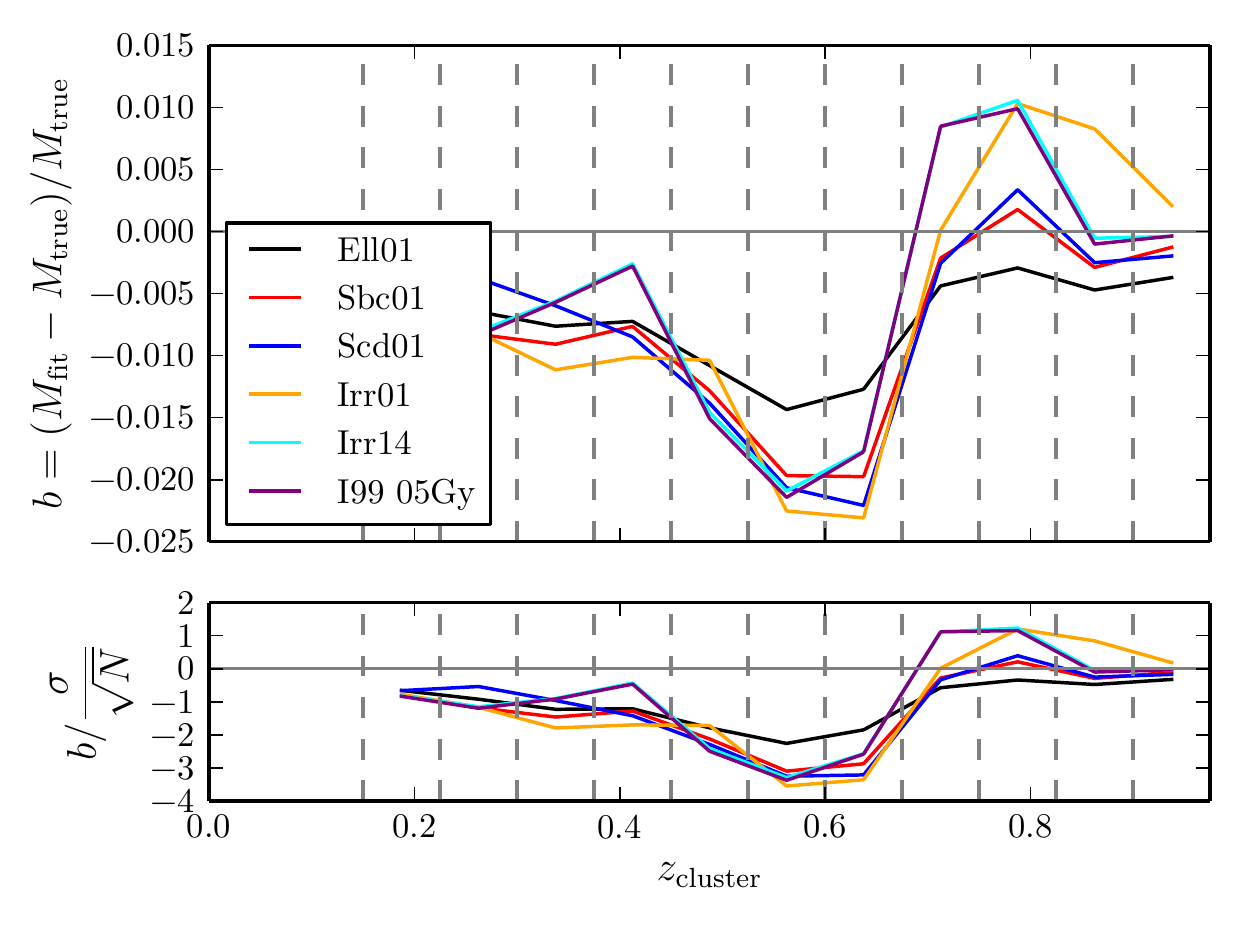}
			\caption{$M_{\mathrm{true}}=0.88 \times 10^{14} \, h^{-1} \, \mathrm{M_\odot}$ \\ $ 0.2 \, R_{\mathrm{vir}} \leq r \leq 1 \, R_{\mathrm{vir}}$}
		\end{subfigure}
        \begin{subfigure}{0.49\textwidth}
           	\centering
			\includegraphics[width=\textwidth]{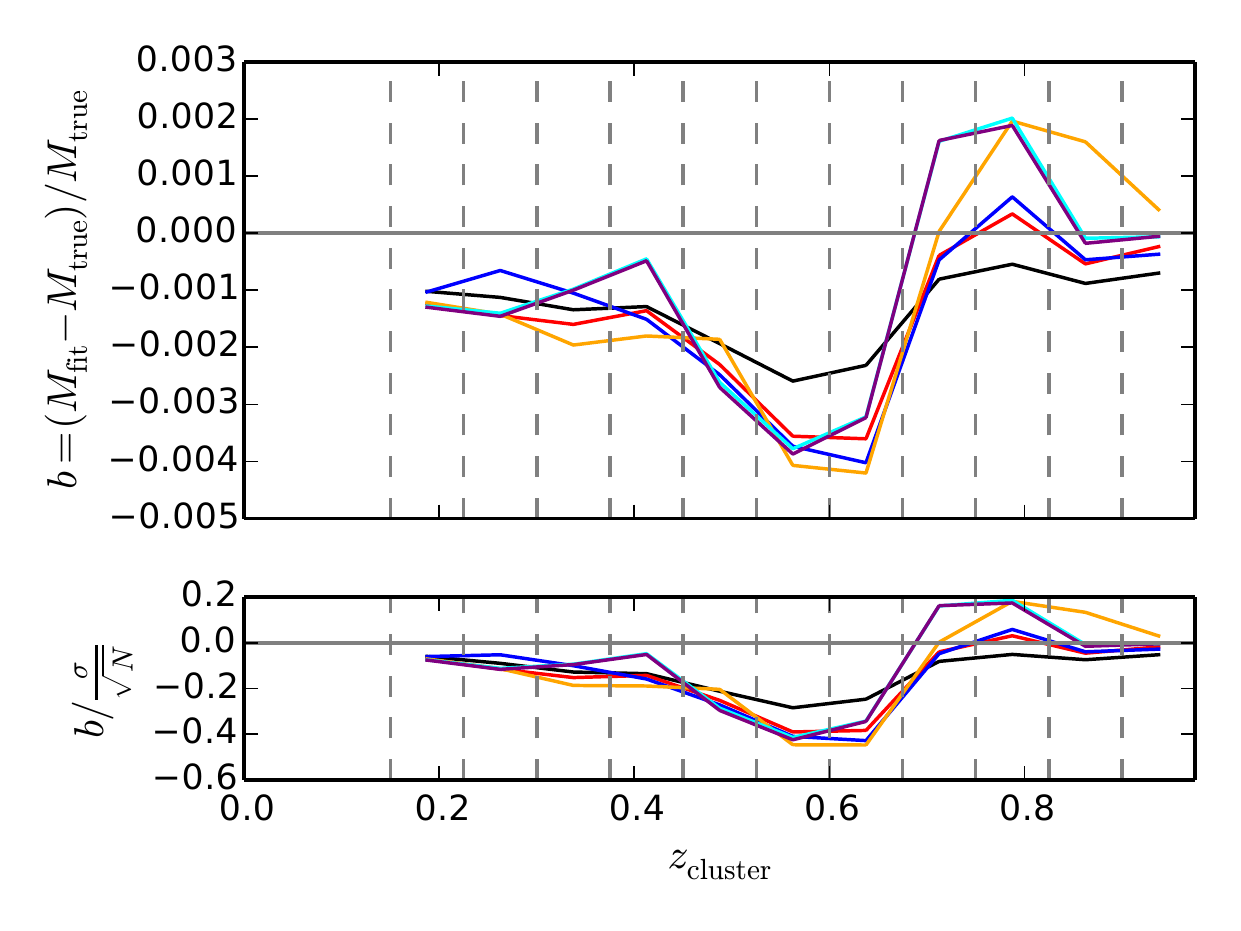}
			\caption{$M_{\mathrm{true}}=0.88 \times 10^{14} \, h^{-1} \, \mathrm{M_\odot}$ \\ $ 1 \, R_{\mathrm{vir}} \leq r \leq 2 \, R_{\mathrm{vir}}$}
		\end{subfigure} \\
		\begin{subfigure}{0.49\textwidth}
			\centering
			\includegraphics[width=\textwidth]{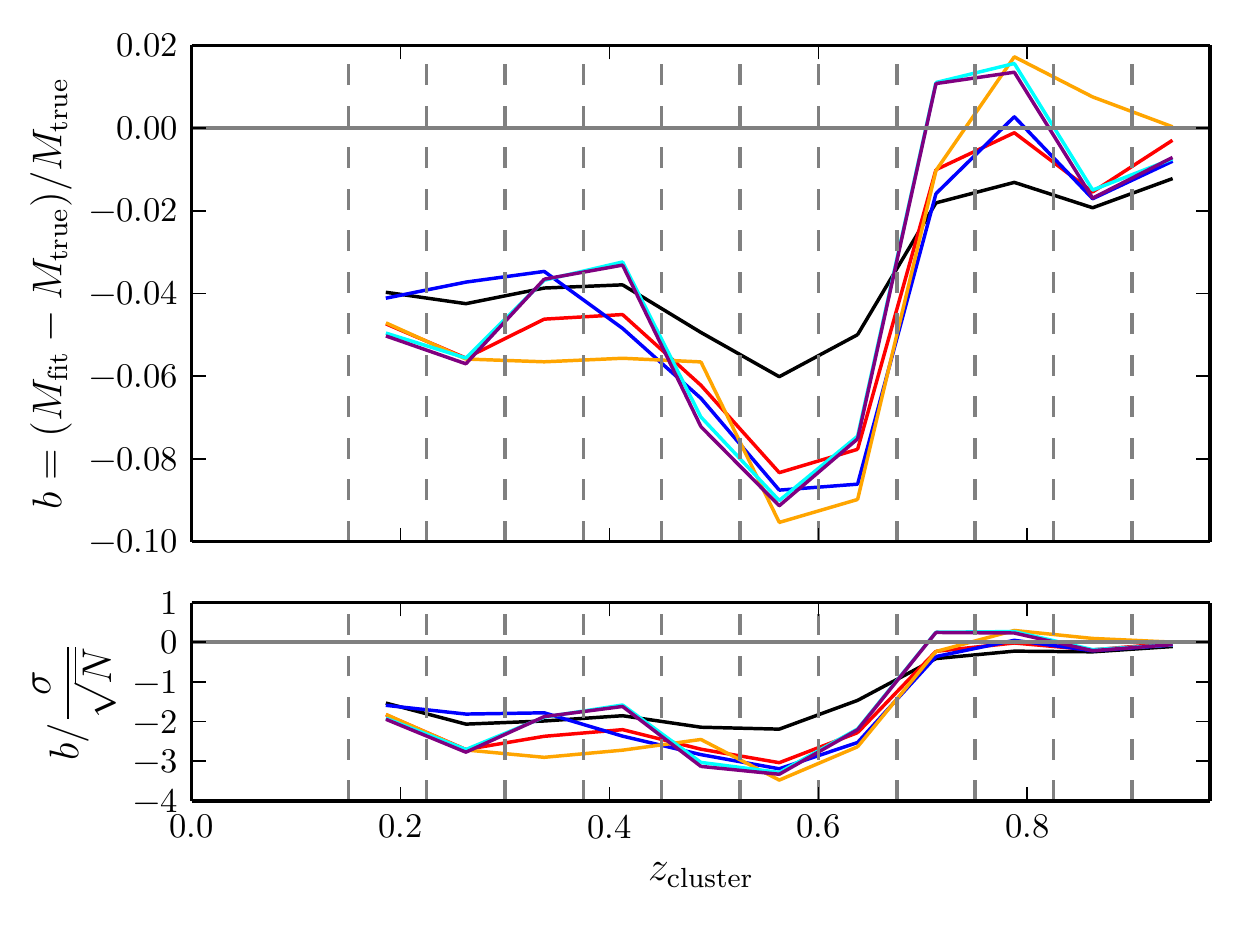}
			\caption{$M_{\mathrm{true}}=8.76 \times 10^{14} \, h^{-1} \, \mathrm{M_\odot}$ \\ $ 0.2 \, R_{\mathrm{vir}} \leq r \leq 1 \, R_{\mathrm{vir}}$}					
		\end{subfigure}
        \begin{subfigure}{0.49\textwidth}
           	\centering
			\includegraphics[width=\textwidth]{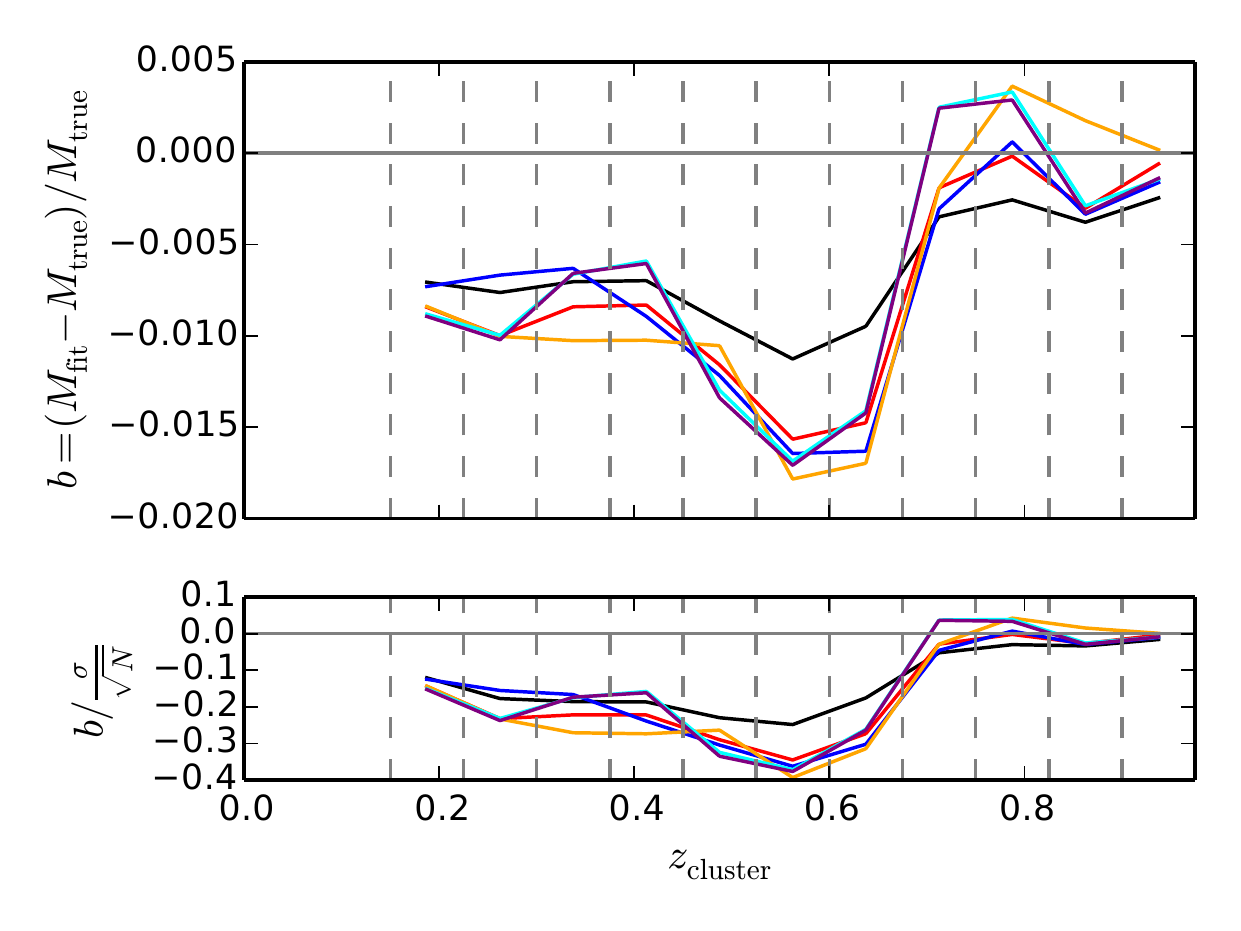}
			\caption{$M_{\mathrm{true}}=8.76 \times 10^{14} \, h^{-1} \, \mathrm{M_\odot}$ \\ $ 1 \, R_{\mathrm{vir}} \leq r \leq 2 \, R_{\mathrm{vir}}$}			
		\end{subfigure}
		\caption{\\ \textbf{Upper panels:} Relative bias $b$ due to imperfect assignment of photometric redshifts of cluster members which thus contaminate the source redshift distribution as a function of true cluster redshift $z_{\mathrm{cluster}}$, SED type (i.e., ''Ell01`` corresponding to elliptical, ''Sbc01`` and ''Scd01`` corresponding to spiral, and various ''Irr`` corresponding to irregular galaxy SED templates), and cluster mass. Panels \textbf{(a)} and \textbf{(c)} show the lowest and highest mass in the first radial bin, whereas panels \textbf{(b)} and \textbf{(d)} show the corresponding masses in the second radial bin.\\
		\textbf{Lower panels:} Comparison of the relative bias $b$ from the upper panels with the corresponding relative uncertainties derived for a stack of galaxy clusters $\sigma / \sqrt{N}$ (cf. Fig.~\ref{fig:forecast} and Table~\ref{tab:numbers}) again as a function of cluster redshift, SED type and cluster mass. Note the different scale of the y-axis between the radial bins.}
		\label{fig:forecast_photoz}
	\end{minipage}
\end{figure*}

We compare the effective lensing efficiencies (cf. equation~\ref{eq:sig_crit_pz}) derived from these two distributions over the fitting range of the first radial bin between $0.2 \, R_{\mathrm{vir}} \leq r \leq 1 \, R_{\mathrm{vir}}$ in Fig.~\ref{fig:sig_crit_pz} for a cluster of mass $M =  8.76 \times 10^{14} \, h^{-1} \, \mathrm{M_\odot}$ at redshift $z = 0.56$. The cluster redshift distribution is assumed to consist entirely of elliptical galaxies (i.e., SED type ``Ell01'') in this case. In the lower panel we show the excess in effective lensing efficiency due to an excess in the cluster redshift distribution based on misidentified photometric redshift estimates. 
In general, applying the foreground cut $p(z \leq z_{\mathrm{cluster}} + 0.15) = 0$ already removes a large fraction of cluster members from the source background distribution. However, there still remains a small fraction of cluster members with wrongly assigned (higher) photometric redshifts which are then erroneously treated as members of the source redshift distribution. This affects the lensing efficiency as shown in the lower panel of Fig.~\ref{fig:sig_crit_pz}. The modified lensing efficiency then leads to a bias in a weak lensing based mass derivation. 
Hence, we study the impact of this effect by repeating our previous weak lensing analysis by replacing the analytic expression from equation~\ref{eq:pz} for $p_{\mathrm{src}}(z)$ with $p_{\mathrm{mod}}(z)$ for two fiducial clusters with masses of $M =  0.88 \times 10^{14} \, h^{-1} \, \mathrm{M_\odot}$ and $M = 8.76 \times 10^{14} \, h^{-1} \, \mathrm{M_\odot}$ at 11 equidistant redshifts between $ 0.1875 \leq z_{\mathrm{cluster}} \leq 0.9375$. We always include large-scale structure contributions in the error budget. Moreover, we consider two radial bins $i$ per cluster, where the first bin spans radii in the range $0.2 \, R_{\mathrm{vir}} \leq r \leq 1 \, R_{\mathrm{vir}}$ and the second bin extends from $1 \, R_{\mathrm{vir}}$ to $2 \, R_{\mathrm{vir}}$. Typical numbers we find for the fraction of cluster members within these bins are $\sim 50 \%$ and $\sim 32 \%$, respectively. The remainder of cluster galaxies is concentrated in the very centre of the cluster, i.e., between $\sim 0 \leq r \leq 0.2 \, R_{\mathrm{vir}}$.

The results of this analysis are summarized in Fig.~\ref{fig:forecast_photoz} and reveal that the impact of cluster members scattered into the lensing source sample is a minor concern for an individual cluster as indicated in the upper panels where we show the relative mass bias, $b$, as a function of true cluster redshift, $z_{\mathrm{clus}}$. When we compare the relative uncertainties for stacks of galaxy clusters, $\sigma /\sqrt{N}$, as obtained in the previous section (cf. Fig.~\ref{fig:forecast} and Table~\ref{tab:numbers}) with the level of expected relative bias $b$ due to imperfect photometric redshift estimates of cluster members as shown in the lower panels of Fig.~\ref{fig:forecast_photoz}, we find that the bias is strongly dependent on the radial bin under consideration, as expected. For the first bin between $0.2 \, R_{\mathrm{vir}}$ and $1 \, R_{\mathrm{vir}}$ the bias can be severe for certain combinations of SED type and redshift. In the second bin between $1 \, R_{\mathrm{vir}}$ to $2 \, R_{\mathrm{vir}}$ the bias is always within the expected statistical uncertainties though for all SED types and redshifts but it can amount up to $~40\%$ of the statistical uncertainties. 

The dependence of the bias on the SED type used for the galaxy templates in the derivation of the photometric redshift estimates is exaggerated for most SED types because we always consider all members to consist of only one SED type. Galaxies with irregular SED types show the strongest bias whereas the lowest bias is found for elliptical SED types although it is then of equal strength as the statistical uncertainties. In real clusters though, the majority of member galaxies will consist of (red) elliptical galaxies with rising fractions of (blue) spiral galaxies as a function of increasing cluster redshift (Butcher--Oemler effect, \citealt{ButcherOemler1984}). Hence, the total bias will be dominated by a combination of the individual biases of these SED types, whereas contributions from irregular galaxies, which create the strongest bias in our analysis, will be much smaller in reality. Furthermore, a mass estimate derived over the full fitting range will be less affected by this bias since such an analysis corresponds to taking the average over both radial bins. The same conclusions will also hold for a measurement of cluster density profiles so that the photometric redshift bias will be negligible in this case, too. 

For high redshift clusters most of the photometric misidentifications will be scattered to redshifts in the foreground of the cluster, especially when considering the imposed detection limits. Thus the ratio of the bias over the statistical uncertainties will flatten towards higher cluster redshifts (cf. lower panels of Fig.~\ref{fig:forecast_photoz}).

\subsection{Miscentring bias}
\label{subsec:misc}
The position of the minimum of the cluster potential, its centre, is unknown in general. The position of the brightest cluster galaxy (BCG) or the centre of the X-ray emission can be used as a tracer for the cluster centre, but they do not have to coincide with the centre as determined by weak lensing, especially not in unrelaxed haloes. Hence, we need to consider a distribution in offsets.

For a two-dimensional offset $r_{s}$ we can calculate the azimuthally averaged convergence profile $\kappa(r)$ \citep{Yang2006}:
\begin{equation}
 \kappa(r \rvert r_{s}) = \frac{1}{2\pi} \int_{0}^{2\pi} \kappa_{\mathrm{NFW}}\left(\sqrt{r^2+r_{s}^2+2r \cdot r_{s}\cdot \cos(\theta)}\right) \diff \theta \, ,
\end{equation} 
where $\kappa_{\mathrm{NFW}}$ is the convergence for a spherically symmetric NFW--profile as given in equation~\ref{eq:kappa}.

\begin{figure}
		\centering
		\includegraphics[width=84mm]{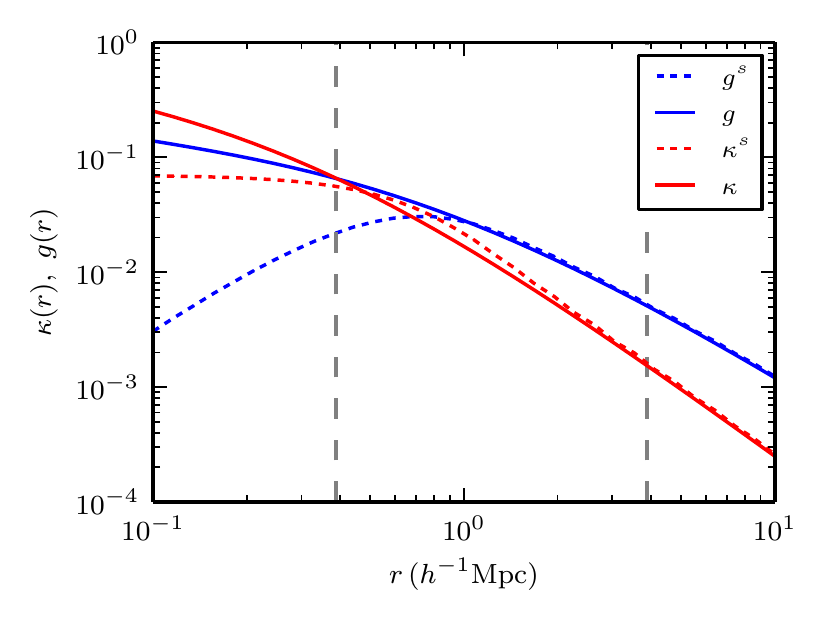}
		\caption{Reduced shear signals (blue) for a perfectly centred ($f=1$) halo (solid line) 
		of mass $M = 8.76 \times 10^{14} \, h^{-1}\mathrm{M_\odot}$ and its maximally 
		miscentred ($f=0$) equivalent (dashed line) with $\sigma_s = 0.42 \, 
		h^{-1}\mathrm{Mpc}$ (see text for details). The convergence is shown in red. The 
		vertical dashed lines indicate the region which we include for fitting in our 
		subsequent analysis ($0.2 \, R_{\mathrm{vir}} \leq r \leq 2 \, R_{\mathrm{vir}}$).}
		\label{fig:diluted_shear}
\end{figure}

Based on SDSS-like mock catalogues for the \texttt{maxBCG} BCG finder algorithm, \citet{Johnston2007} found that the distribution of offsets follows a two-dimensional Gaussian distribution. Although the miscentring in these mock catalogues can only be caused by misidentifications due to the algorithm by construction, we still consider their results to be sufficiently good approximations since in practice one would also employ algorithms to determine miscentring offsets in samples as large as expected to be found in the \textit{Euclid} survey. Therefore, we follow \citet{Johnston2007} by assuming a two-dimensional Gaussian distribution of offsets,  
\begin{equation}
\label{eq:miscentring}
 P(r_{s}) = \frac{r_{s}}{\sigma_{s}^2} \exp(-0.5(r_{s}/\sigma_{s})^2) \, ,
\end{equation}
where the effective scale length was typically found to be $\sigma_{s} = 0.42 \, h^{-1} \mathrm{Mpc}$ independent of cluster richness.
The resulting convergence for miscentred clusters is then a convolution of the above equations which yields
\begin{equation} 
 \kappa^s(r) = \int P(r_{s}) \, \kappa(r \rvert r_{s}) \diff r_{s} \, .
\end{equation}
From this convolved convergence we can derive the reduced shear $g^s$ for a miscentred cluster and employ the lensing formalism of Section~\ref{subsec:chi2}. The total shear signal for a stack of miscentred clusters must be weighted by the fraction $f$ of correctly centred clusters though, i.e.,
\begin{equation}
g^{\mathrm{tot}} = f\cdot g + (1-f)\cdot g^s \, .
\end{equation} 
In Fig.~\ref{fig:diluted_shear} we show both the reduced shear signal and the convergence of a maximally miscentred halo ($f=0$ and $\sigma_{s} = 0.42 \, h^{-1}\mathrm{Mpc}$) and compare it to the expected signal of a perfectly centred halo of the same mass. The effect of miscentring is to dilute the shear signal in the inner regions of the cluster. Hence ignoring this will bias masses low with a dependence on the innermost radius used while fitting. 

\begin{figure*}
	\centering
	\begin{minipage}{180mm}
    	\centering
		\begin{subfigure}{0.49\textwidth}
			\centering			
			\includegraphics[width=\textwidth]{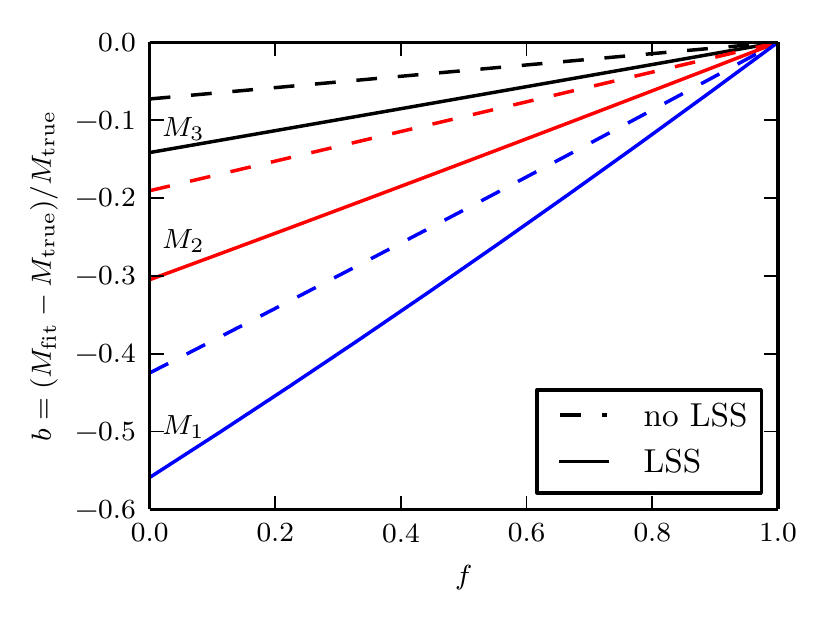}
			\caption{}
			\label{fig:bias_f}
		\end{subfigure}
        \begin{subfigure}{0.49\textwidth}
           	\centering
			\includegraphics[width=\textwidth]{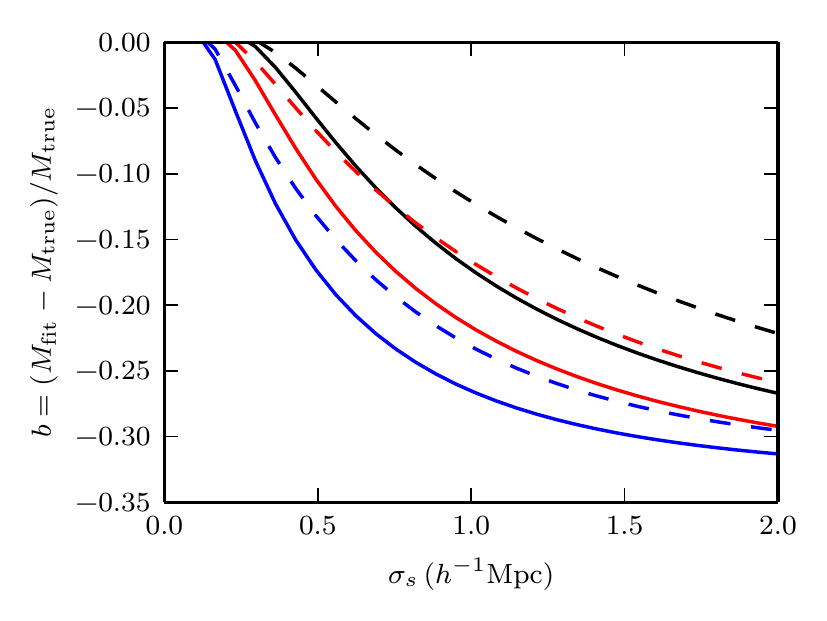}
			\caption{}
			\label{fig:bias_sigma}
		\end{subfigure}
		\caption{\\ \textbf{(a)} Bias in halo mass introduced by miscentring as a function of the 
		fraction of centred haloes $f$ (for a fixed width $\sigma_{s} = 0.42 \, 
		h^{-1}\mathrm{Mpc}$ of the miscentring distribution) again with (solid lines) and 
		without	(dashed lines) large-scale structure contribution. The bias is shown for halo masses of $M_1 = 8.76 \times 10^{13} \, h^{-1}\mathrm{M_\odot}$ (blue), $M_2 = 3.50 
		\times 10^{14} \, h^{-1}\mathrm{M_\odot}$ (red) and $M_3 = 8.76 \times 10^{14} \, 
		h^{-1}\mathrm{M_\odot}$ (black). \\
		\textbf{(b)} Bias in halo mass as a function of the width $\sigma_s$ of the 
		miscentring distribution for a fixed fraction of centred haloes of $f = 0.75$ again 
		with (solid lines) and without (dashed lines) large-scale structure contribution for 
		the same halo masses as in (a).}
	\end{minipage}
\end{figure*}

Before considering strategies for mitigating the bias, we first examine its size further in Fig.~\ref{fig:bias_f} and Fig.~\ref{fig:bias_sigma}. Here, the miscentred shear signal is parametrized by the mass of the halo $M$ and the two miscentring parameters.
In Fig.~\ref{fig:bias_f} we fit a fiducial, centred cluster signal to the signal of a miscentred halo with a varying fraction of centred haloes ($0 \leq f \leq 1$) for three fiducial masses in the range $0.87 \leq M/ (10^{14} \, h^{-1}\mathrm{M_\odot}) \leq  8.76$ (dashed lines). The width of the miscentring distribution is kept fixed at $\sigma_s= 0.42 \, h^{-1} \mathrm{Mpc}$. Miscentring introduces a bias in the recovered mass and increases with decreasing fraction of centred haloes $f$. Furthermore, the bias is dependent on the fiducial halo mass and increases from high halo masses (of order $~10^{15}\, h^{-1}\mathrm{M_\odot}$) to low halo masses (of order $~10^{14}\, h^{-1}\mathrm{M_\odot}$). The dependence of the bias on the fiducial halo mass (for fixed $\sigma_s$) is caused by that the dilution of the shear signal due to miscentring is higher for low mass haloes.
Taking then again large-scale structure into account increases the mass bias even further (blue, solid lines) because cosmic noise reduces the relative contribution of large scales to the shear signal.  
\begin{table*}
\begin{minipage}{126mm}
\caption{Requirements on the precision of the fraction of centred haloes $f$ and the relative error $\Delta \sigma_s/\sigma_s$ of the width $\sigma_s$ of the miscentring distribution.}
\label{tab:constraints_f_sigma}
\begin{tabular}{@{}ccccc}
\hline
  $M_{\mathrm{true}} \, (\times 10^{14} \, h^{-1} \mathrm{M_\odot})$ &
  $\Delta f_{\mathrm{max}}$ &
  $\left(\frac{\Delta \sigma_s}{\sigma_s}\right)_{\mathrm{max}} \, (\sigma_s=0.42 \, h^{-1} \mathrm{Mpc})$ &
  $\left(\frac{\Delta \sigma_s}{\sigma_s}\right)_{\mathrm{max}} \, (\sigma_s=1 \, h^{-1} \mathrm{Mpc})$ \\
\hline
  0.876 & 0.0060 & 0.019 & 0.032 \\
  3.500 & 0.0171 & 0.032 & 0.036 \\
  8.763 & 0.0747 & 0.088 & 0.064 \\
\hline
\end{tabular}

\medskip
We refer the reader to the text for details on the calculation of these requirements.
\end{minipage}
\end{table*}

We repeat these calculations, but this time we keep the fraction of centred haloes fixed at $f = 0.75$ and leave the width of the miscentring distribution, $\sigma_s$, free to vary. The results are shown in Fig.~\ref{fig:bias_sigma} for the same fiducial masses. The functional behaviour is this time more complex, but again it is apparent that including large-scale structure contributions (solid lines) also increases the bias caused by miscentring.

Thus, the combined effect of large-scale structure and miscentring is to increase the uncertainties in the determination of the halo mass and to introduce a mass bias (towards lower masses). 
By comparing the findings of Fig. \ref{fig:bias_f} and Fig. \ref{fig:bias_sigma} to the relative uncertainties on the cluster mass as provided in Table~\ref{tab:numbers}, we conclude that for the mass estimate of a single cluster the bias due to miscentring is negligible compared to the relative uncertainty $\sigma$. For a stacked signal with statistics such as will be provided by the \textit{Euclid} survey, this is no longer the case, since the statistical error $\sigma$ will be reduced by a factor $1/\sqrt{N}$ (cf. Table~\ref{tab:numbers}). 
Hence, it will be necessary to determine the miscentring parameters $f$ and $\sigma_s$ to within a few per cent in order to derive accurate mass determinations. 

In order to quantify this, we calculate upper limits for the precision to within which $f$ and the relative error on $\sigma_s$ must be known such that the bias due to miscentring is smaller than the $1\sigma$-error in mass (cf. Table~\ref{tab:numbers}). The upper bound on the precision of $f$ is thus defined as 
\begin{equation}
\Delta f_{\mathrm{max}} = \frac{\sigma}{\sqrt{N}} \left(\frac{\mathrm{d}\Delta M}{\mathrm{d}f}\right)^{-1} \, . 
\end{equation}
Furthermore, the upper bound on the relative error $\Delta \sigma_s/\sigma_s$ of the width $\sigma_s$ is given as 
\begin{equation}
\left(\frac{\Delta \sigma_s}{\sigma_s}\right)_{\mathrm{max}} = \frac{\sigma}{\sqrt{N}} \left(\left.\frac{\mathrm{d}\Delta M}{\mathrm{d}\sigma_s}\right|_{\sigma_s}\right)^{-1} \, .
\end{equation}
The values corresponding to the three fiducial masses $M_{\mathrm{true}}$ of Fig.~\ref{fig:bias_sigma} including large-scale structure contributions are provided in Table~\ref{tab:constraints_f_sigma}.   

Next, we explore if these conclusions also hold if we marginalize over the miscentring parameters expressing our lack of knowledge about them. Note as well, that we still assume equation~\ref{eq:miscentring} to hold which in itself is an important assumption in this regard.
Hence, we explore the parameter space by fitting a miscentred shear signal to the shear signal of a fiducial model with mass $M = 8.75 \times 10^{14} \, h^{-1}\mathrm{M_\odot}$, $f=1$ at redshift $z = 0.1875$ for different values of $M$, $\sigma_s$, and $f$. All three parameters are varied for $50$ different masses in the range $2.03 \times 10^{14} \, h^{-1}\mathrm{M_\odot} \leq M \leq 2.03 \times 10^{15} \, h^{-1}\mathrm{M_\odot}$, $40$ different values of $\sigma_s$ in the range $0 \leq \sigma_s \leq 2 \, h^{-1}\mathrm{Mpc}$, and $20$ different values of the fraction of centred haloes in the range $0 \leq f \leq 1$, respectively.    

For the fitting we adopt again the formalism of Section~\ref{subsec:chi2} always including large-scale structure contributions.
We have repeated these calculations for a second halo with fiducial mass $M=3.50 \times 10^{14} \, h^{-1}\mathrm{M_\odot}$ (where we adjust the mass ranges accordingly). 

\begin{figure*}
	\centering
	\begin{minipage}{180mm}
    	\centering
		\begin{subfigure}{0.49\textwidth}
			\centering			
			\includegraphics[width=\textwidth]{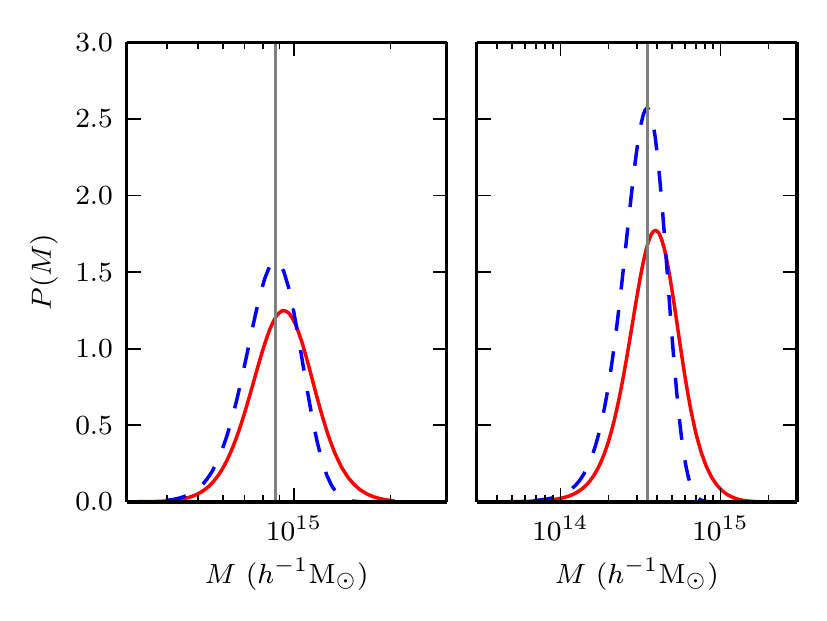}
			\caption{}
			\label{fig:comparison_marg_flat}
		\end{subfigure}
        \begin{subfigure}{0.49\textwidth}
           	\centering
			\includegraphics[width=\textwidth]{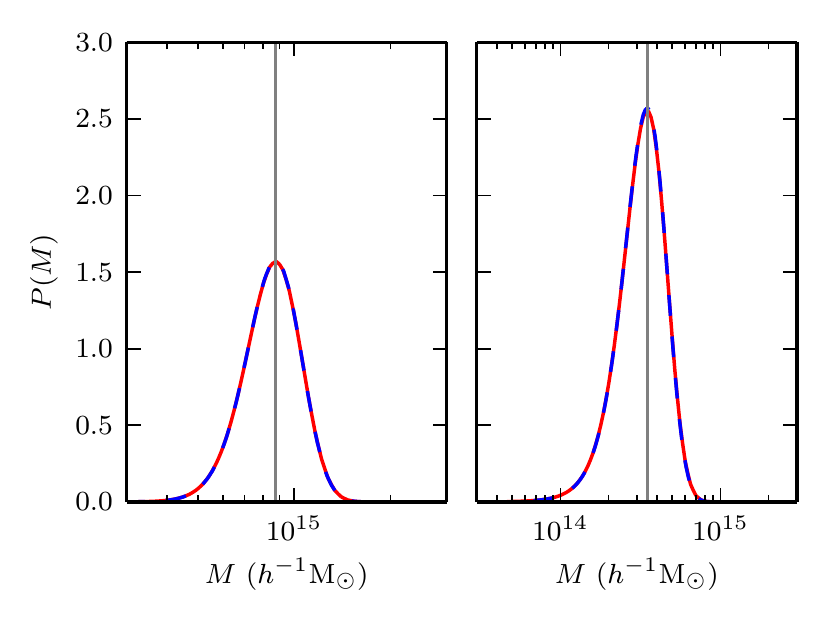}
			\caption{}
			\label{fig:comparison_marg_gauss}
		\end{subfigure}
		\caption{Comparison of likelihoods for two haloes of masses $M = 8.75 \times 10^{14} \, 
		h^{-1}\mathrm{M_\odot}$ (left) and $M = 3.50 \times 10^{14} \, h^{-1}\mathrm{M_\odot}$ 
		(right), respectively (blue, dashed lines) and likelihoods for the same mass haloes but 
		marginalized over miscentring parameters $\sigma_s$ and $f$ (red, solid lines) assuming the following priors for both parameters: \\
		\textbf{(a)} Flat priors. \\
		\textbf{(b)} Gaussian priors.\\
		The grey, solid line shows the fiducial mass for each halo and we always account for large-scale structure contributions in the error budget. All haloes are at redshift $z = 0.1875$.}
	\end{minipage}
\end{figure*}

In Fig.~\ref{fig:comparison_marg_flat} we compare the marginalized probability distributions assuming flat priors on $\sigma_s$ and $f$ over the ranges indicated above for the two haloes with their perfect fit counterparts (cf. Fig.~\ref{fig:PDF_mass}). The marginalization over $f$ and $\sigma_s$, respectively, introduces a small bias due to the truncation of the prior on $\sigma_s$ at $2 \, h^{-1}\mathrm{Mpc}$. However, the assumption of flat priors on both miscentring parameters is too pessimistic.

Already with current data it is possible, particularly for the most massive clusters, to derive more realistic priors for the miscentring parameters than flat ones. Based on measurements of the displacement between the BCG and the maximum of the X-ray radiation from the intracluster medium from a sample of 53 massive galaxy clusters presented in \citet{Bildfell2008}, we have estimated errors on $f$ and $\sigma_s$ by bootstrapping the distribution of measured off-centre radii. We find $\Delta f = 0.04$ and $\Delta \sigma_s = 0.01 \, h^{-1} \mathrm{Mpc}$ independent of the number of bootstraps once this exceeds $\sim 1000$. Interpreting these errors then as widths of Gaussian priors on $f$ and $\sigma_s$, respectively, yields qualitatively unbiased mass estimates when marginalizing over these priors instead of flat priors, as shown in Fig.~\ref{fig:comparison_marg_gauss}. 

In the very near future a mapping of the X-ray sky will be carried out by eROSITA \citep{Merloni2012} which is also aimed at detecting a large sample ($\sim 10^5$) of galaxy clusters. This sample can then be used to derive even better priors on the miscentring parameters which will mitigate the effect of the miscentring bias for \textit{Euclid} weak lensing cluster masses entirely. Moreover, the combination of the eROSITA and \textit{Euclid} cluster surveys will also allow to reduce the uncertainty on the hydrostatic mass bias between X-ray and weak lensing mass estimates so far that determining the sum of neutrino masses will be possible with cluster counts (cf. figure~12 from \citealt{Planck2015a}).    


\section{Conclusions}
\label{sec:conc}
We have investigated the level of statistical uncertainties and scrutinized various sources of systematic errors in the determination of masses for stacks of galaxy clusters employing weak gravitational lensing. Throughout the analysis we have been focussing on future large area surveys, in particular the \textit{Euclid} survey, so that the predominant source of statistical uncertainty, i.e., shape noise, is as small as possible when referring to (large) stacks of galaxy clusters. 

In Section~\ref{sec:stat_uncert} we have established the level of expected statistical uncertainties on the mass of (stacks of) galaxy clusters. We emphasize that contributions of cosmic noise must be included in realistic statistical uncertainties. However, the level of expected statistical uncertainties is very low (cf. Table~\ref{tab:numbers} and Fig.~\ref{fig:forecast}). Thus, at this level of expected statistical precision sources of systematic errors become important which are still negligible compared to shape noise in current and ongoing surveys. Furthermore, the \textit{Euclid} survey poses an upper limit of $m < 2 \times 10^{-3}$ \citep{Laureijs2011} on one of the predominant biases of weak lensing, i.e., the multiplicative bias $m$. The other predominant bias, i.e., the additive bias, is negligible when referring explicitly to stacks of clusters due to the averaging process involved in that kind of analysis. Hence, we expanded our analysis to other sources of bias instead.

We have addressed one of these additional biases in Section~\ref{subsec:photo-z} in which we investigated the effect on the accuracy of lensing masses when accounting for imperfect photometric redshift assignment to cluster member galaxies. Typically, a small fraction of these will be scattered to higher redshifts and thus mimic background source galaxies. This leads to a decreased weak lensing signal and thus requires the assignment of higher cluster masses during the fitting. We have shown that this bias is significant, especially for analyses using radii between $0.2 \, R_{\mathrm{vir}}$ and $1 \, R_{\mathrm{vir}}$. However, even including larger radii out to $2 \, R_{\mathrm{vir}}$ will still require to properly account for this bias in the full analysis and we strongly recommend to study the impact of this effect further in the context of more detailed simulations.

Another source of systematic error is the effect of miscentring, i.e., the ambiguity in the choice of an observational cluster centre. In general, a displacement of the observed cluster centre and the true cluster centre leads again to a dilution of the shear signal and thus to higher mass estimates. Already with currently available data it is possible though to derive realistic priors for the miscentring parameters so that it will be possible to mitigate this bias entirely even in the case of \textit{Euclid} when taking complementary missions such as eROSITA into account.

Finally, we want to emphasize that our analysis of these additional sources of systematic errors are all based on (simple) analytic models. Eventually, these will have to be reassessed by extensive numerical simulations in order to derive more realistic bounds and quantitative estimates. Our analysis, however, is meant to point to the relative importance and order of magnitude predictions of these systematic errors in order to supply the community with a guideline for these future simulations. 

\section*{Acknowledgements}
We thank the anonymous referee for very detailed comments which helped to further improve this work and its presentation.
FK acknowledges support from a de Sitter Fellowship of the Netherlands Organization for Scientific Research (NWO). HH and ME acknowledge support from the European Research Council under FP7 grant number 279396.\\
Plots in this paper were produced with \texttt{Python} and its \texttt{matplotlib} \citep{Hunter2007}. Cosmology related calculations were performed using the \texttt{Python} package \texttt{CosmoloPy}\footnote{\url{http://roban.github.com/CosmoloPy/}}.

\bibliographystyle{mn2e}
\bibliography{bibliography}

\label{lastpage}
\end{document}